\documentclass[journal,10pt,twocolumn]{IEEEtran}

\usepackage{cite}
\usepackage{amsmath,amssymb,amsfonts}
\usepackage{algorithmic}
\usepackage{graphicx}
\usepackage{textcomp}
\usepackage{xcolor}
\usepackage{CJKutf8}
\usepackage{url}
\usepackage{tikz}
\usetikzlibrary{arrows.meta,positioning,fit,calc}
\usepackage{rotating} 
\usepackage{tabularx}
\usepackage{booktabs} %

\def\BibTeX{{\rm B\kern-.05em{\sc i\kern-.025em b}\kern-.08em
    T\kern-.1667em\lower.7ex\hbox{E}\kern-.125emX}}
\begin{document}

\title{\color{black} Generalized green and red concepts for  artificial intelligence, big data, and information and communication technologies: Multidimensional framework and countermeasures}
\author{Jinsong Wu, {\em Senior Member, IEEE}, Yixin Xiao, Haowei Bi, and Huamin Fang 
\thanks{Corresponding author: Jinsong Wu (email address: wujs@ieee.org).\\
Jinsong Wu is with School of Artificial Intelligence, Guilin University of Electronic Technology, Guilin, China, and Department of Electrical Engineering, University of Chile, Santiago, Chile, email address:  wujs@ieee.org. \\
Yixin Xiao, Haowei Bi, and Huamin Fang are with  School of Artificial Intelligence, Guilin University of Electronic Technology, Guilin, China.
 }}
\maketitle

\begin{abstract}
\color{black} The rapid expansions of artificial intelligence (AI), big data (BD), and information and communication technologies (ICT) are driving substantial growth in electricity, water, and material demands, with data centers projected to consume 6.7--12\% of U.S. electricity by 2028 (up from 4.4\% in 2023) and AI-related water demand estimated at 4--6 billion m$^3$ annually by 2027. Existing ``green AI'' and ``red AI'' discussions have largely emphasized computational efficiency and carbon emissions, providing limited guidance for lifecycle trade-offs, service-versus-use impacts, and broader sustainability externalities beyond energy. To bridge this gap, this paper proposes generalized green AI (GGAI) and generalized red AI (GRAI) and extends the taxonomy to red BD (RBD) and red ICT (RICT) as well as red electronic and electrical engineering (REEE) \& red digital technologies (RDT) under a unified Environmental-Economic-Social-Technical/Operational (E/Ec/S/T) framework. Based on representative footprint indicators in the large-model era (electricity, carbon, water, and e-waste), we analyze how risks propagate across AI-BD-ICT stacks and across service-use boundaries. We further organize actionable countermeasures into a four-tier taxonomy aligned with the Environmental-Economic-Social-Technical/Operational framework: metric-level reform for accountability, revision of information and communication technologies design principles and demand shaping, efficiency-oriented LLM development practices (data, model, training, and scaling), and energy-infrastructure transition. The proposed framework operationalizes the Environmental-Economic-Social-Technical/Operational-aligned assessment and intervention prioritization across the AI-BD-ICT stacks, enabling sustainability-by-design without sacrificing utility.
\end{abstract}


\section{Introduction}\label{sec_intro}
{
\color{black} 
Climate change has become one of the most critical global challenges. If warming cannot be slowed down, each additional increase will intensify heat extremes, floods and droughts, sea-level rise, and associated health and economic damage, while also increasing the risks of irreversible ecosystem loss and adaptation limits, as concluded by IPCC AR6 \cite{IPCC_AR6_SYR_2023} and corroborated by recent global climate and health reports \cite{wmo2024climate,JL_Romanello2024-kk}. In high-emission scenarios, some regions may face heat conditions that exceed or exceed human survivability thresholds for substantial fractions of a year; in extreme cases, climate change may trigger cascading social stresses \cite{doi:10.1073/pnas.2305427120,JCCH_BUTLER2024100329}.

Alongside traditional high-emission sectors, the rapid expansion of artificial intelligence (AI), big data, and information and communication technologies (ICT) is emerging as a non-negligible driver of electricity demand, water withdrawal, material consumption, and broader socio-technical externalities. Although these technologies enable transformative benefits, their energy-intensive infrastructures and accelerated deployments raise concerns about long-term sustainability and the distribution of environmental burdens \cite{JJ_deVries2023,doi:10.1126/science.adt5536,JN_Istrate2024,doi:10.1177/0958305X221118877}. Previous studies and policy discussions on green development have focused largely on conventional industries, whereas existing technology-oriented frameworks often emphasize computational efficiency and carbon emissions, leaving broader sustainability dimensions inadequately structured for AI/BD/ICT systems \cite{wu2018ICTSDG,J_ISJ_GBD_wu2016big,FREITAG2021100340,JHSSC_Wang2023}. To contextualize these impacts quantitatively, Table~I summarizes representative indicators related to footprints commonly used to characterize AI/BD/ICT sustainability concerns.

This paper addresses the above gaps by establishing a unified and operational sustainability viewpoint that (i) treats artificial intelligence, big data, and information and communication technologies as a coupled system stack, and (ii) distinguishes between sustainability impacts arising from providing services versus using services, under a multidimensional Environmental (E), Economic (Ec), Social (S), and Technical/Operational (T) perspective. In particular, this work moves beyond energy-centric interpretations of ``green AI'' to a generalized framework that can classify both beneficial and harmful development modes across the full lifecycle of data, models, and infrastructures.

This paper seeks to bridge this gap with the following contributions.
\begin{enumerate}
    \item This paper analyzes possible planetary-scale sustainability risks associated with the development of AI, including energy consumption and carbon footprint, water consumption and vapor-related concerns, e-waste accumulation, and socio-technical externalities in Section~\ref{sec_crisis_AI}.

    \item This paper proposes the concepts of generalized green AI and generalized red AI in Section~\ref{sec_GGAI_GRAI}. It further proposes the concepts of red big data and red information and communication technologies in Section~\ref{sec_BD_ICT_RG}. These classifications will be developed within a unified Environmental (E), Economic (Ec), Social (S), and Technical/Operational (T) framework, extending beyond traditional energy/computation-centric green AI definitions.

    \item This paper analyzes current and future risks and provides actionable technical and governance countermeasures to generalized red AI, red big data, and red ICT in the era of large language models and generative AI in Section~\ref{sec_cr_Red}.
\end{enumerate}

The remainder of this paper is organized as follows: Section~\ref{sec_crisis_AI} analyzes sustainability risks in the large-model era, Sections~\ref{sec_GGAI_GRAI}-\ref{sec_BD_ICT_RG} formalize the generalized Environmental-Economic-Social-Technical/Operational framework for AI/BD/ICT, and Section~\ref{sec_cr_Red} discusses risk trajectories and countermeasure pathways.
}

\section{Planetary sustainability risks induced by {AI}}\label{sec_crisis_AI}
 The dual natures of AI as both a tool of progresses and potential environmental hazards underscore the demands for sustainable
development and the deployment of technologies.

{\color{black} 
\subsection{Overview and representative footprint indicators}
The sustainability risks of modern \mbox{AI} are not limited to carbon emissions. They arise from the coupling of 
compute demand, cooling water, material throughput, and the socio-technical externalities that scale with deployment. 
Because these impacts are highly heterogeneous across regions (grid mix, climate, and water scarcity) and across 
development stages (training vs.\ inference), we summarize representative indicators in Table~\ref{tab_ai_footprints} 
to ground the discussion and avoid over-generalization.

\begin{table*}[tp]
\caption{Representative footprint indicators for \mbox{AI}/LLM development (illustrative; varies by model, site, and 
grid).}
\label{tab_ai_footprints}
\centering
\begin{tabular}{p{0.18\linewidth} p{0.33\linewidth} p{0.34\linewidth}}
\hline
\textbf{Category} & \textbf{Indicator} & \textbf{Representative magnitude (example)} \\
\hline
Energy \& carbon & Data center electricity share (USA) & 4.4\% (2023) projected to 6.7--12\% by 2028 \cite{LBNL_shehabi2024} \\
Energy \& carbon & Training-related emissions (LLM example) & One GPT-3 training run reported $\sim$1.21M lb $CO_2$ emissions \cite{chen2023survey} \\
Water & Global \mbox{AI}-related water demand & 4--6 billion m$^3$ per year projected by 2027 \cite{li2023making} \\
Water & On-site cooling efficiency (WUE) & Typical on-site WUE $\sim$0.2--1.8 L/kWh (site/climate dependent) \cite{ACM_li2025making} \\
Materials \& waste & \mbox{AI}-driven e-waste growth (scenario) & $\sim$16 million tons by 2030 (model-based estimate) \cite{wang2024waste} \\
\hline
\end{tabular}
\end{table*}

\subsection{Energy demand and carbon emissions}
Empirical and cross-country evidence suggests that \mbox{AI} activity can increase $CO_2$ emissions, with stronger 
effects in settings where governance and energy policy are weaker\cite{JEM_ALNAFRAH2025126813}. A key driver is the 
rapid expansion of data-center capacity to support training and serving large-scale models. In the United States, data 
centers are projected to consume 6.7\%--12\% of national electricity by 2028 (up from 4.4\% in 2023)\cite{LBNL_shehabi2024}. 
At the model level, training and tuning large neural networks can carry substantial carbon footprints\cite{CACL_strubell-etal-2019-energy}. 
For example, a single GPT-3 training session has been reported to emit on the order of 1.21 million pounds of $CO_2$e\cite{chen2023survey}. 
OpenAI have not disclosed the resource footprints for GPT-4 and GPT-5 in trainings or inferences, but recent analysis has estimated GPT-4 about $2 \times 10^{25}$ floating point operations per second (FLOPs) and GPT-5 about $7 \times 10^{25}$ FLOPs, whose scale implies multi-GWh electricity even on modern accelerators\cite{epochai2025notesgpt5}.
These figures motivate a 
shift from ``performance-only'' scaling to energy-aware development and deployment strategies.

\subsection{Water use and water-vapor externality}
\mbox{AI} workloads consume water both directly (on-site cooling) and indirectly (electricity generation and upstream 
manufacturing). Large-scale model training can require substantial cooling water, and semiconductor fabrication also demands ``ultrapure water''\cite{su70811260,shen2018chemical}. At the system level, projections suggest that global 
\mbox{AI}-related water demand could reach 4-6 billion m$^3$ annually by 2027\cite{li2023making}. The impacts of water are especially significant in water-stressed basins because evaporative cooling turns withdrawals into consumptive losses; the effectiveness of water use on-site (WUE) can range from $\sim$0.2-1.8 L/kWh depending on climate and cooling choices\cite{ACM_li2025making}. Google reported that its data centers consumed 6.1 billion gallons of water for on-site cooling in 2023 (about 23.1 billion  liters), and an independent synthesis of the same disclosure notes that Google self-owned centers withdrew about 29 billion  liters while consuming (evaporating) larger than 23 billion  liters, which underscored that withdrawals exceed consumptive losses and that local cooling choices strongly shape basin impacts\cite{ACM_li2025making}. Li et al. estimated that the run directly evaporated approximately 0.7 million liters on-site (scope-1) and about 5.4 million liters total when including power-sector water (scope-2). 

Beyond local resource stress, evaporated cooling water returns to the atmosphere as water vapor, a natural but 
radiatively active greenhouse gas\cite{nelson2023effects,stephens1993water}. Although the climate impact of data-center 
water vapor is context-dependent and generally far smaller than global forcing $CO_2$, it remains a relevant externality in regions where both heat rejection and water scarcity are binding constraints.  The emissions of water vapor as a greenhouse gas are less known\cite{J_OLAR_patel2023increase}, which has a relatively short lifespan in the atmosphere and is a natural component of the climate system\cite{nelson2023effects}. As the temperature of the Earth increases, the concentration of water vapor in the air increases because warmer air can hold more moisture\cite{inamdar1998tropical}. The increased water vapor can absorb more infrared radiation, leading to a further rise in the surface temperature of the Earth. Climate models have indicated that water vapor feedback is a crucial positive feedback process, nearly double the sensitivity of surface temperature to carbon dioxide and potentially amplifying it further when interacting with other feedbacks\cite{stephens1993water}.

However, the consequences behind such data could be risks. Unlike on Earth, the atmosphere of Venus is primarily composed of carbon dioxide and surface temperatures and pressures are much higher than those on Earth\cite{svedhem2007venus}. Therefore, unbridled AI development could significantly increase water vapor in the air\cite{pierrehumbert2023runaway}, thus increasing vapor risks. 

\subsection{Hardware turnover and electronic waste}
\mbox{AI} acceleration relies on rapid upgrades of GPUs/TPUs and supporting infrastructure, which can increase 
embodied emissions and electronic waste (e-waste). Scenario-based material-flow analyses estimate that \mbox{AI}-related 
hardware turnover could contribute to e-waste growth toward the end of this decade\cite{wang2024waste}. E-waste contains 
hazardous substances (e.g., heavy metals and organic compounds) and can pose risks to ecosystems and human health if 
collection, recycling, and disposal are inadequate\cite{bhoi2014waste,heacock2016waste}. Consequently, sustainability 
strategies for \mbox{AI} should include lifecycle measures such as repairability, longer refresh cycles, and verifiable 
recycling pipelines, alongside compute-efficient model design\cite{fraga2021green}.

\subsection{Other sustainability and socio-technical externalities}
In addition to environmental footprints, large-scale \mbox{AI} may introduce broader sustainability risks that interact 
with social and economic systems:
\begin{enumerate}
\item \textit{Resource allocation and equity:} high capital and energy requirements can concentrate capabilities in a 
small number of regions and firms, potentially widening disparities.
\item \textit{Bias, discrimination, and safety:} deployment of generative models can reproduce biases and harmful content 
if not systematically assessed and mitigated\cite{kasneci2023chatgpt,weidinger2022taxonomy}.
\item \textit{Misuse and misinformation:} realistic synthetic content can be exploited for manipulation and other harmful 
purposes; responsible governance is therefore essential\cite{shin2023human}.
\item Human-machine interaction concerns: ChatGPT could misplace human confidence\cite{chen2023survey}, leading to conflict. From a human-machine conflict perspective, the use of ChatGPT could lead to its substitution for humans and could involve the use of offensive language, causing harm\cite{gill2023chatgpt}.
\end{enumerate}
Overall, these risks reinforce the need to treat sustainability in \mbox{AI} as a multi-dimensional objective rather 
than a single carbon metric.
}

\section{Generalized green {AI} and generalized red {AI}}\label{sec_GGAI_GRAI}

{
\color{black} 
\subsection{Background: red {AI} and green {AI}}\label{subsec_GAI_RAI}
Following \cite{schwartz2020green}, prior work distinguished \emph{red {AI}} ({RAI}) and \emph{green {AI}} ({GAI}) mainly from the viewpoint of computing and energy efficiency:
\begin{enumerate}
    \item \textbf{Red AI} refers to a research approach in artificial intelligence that enhances model accuracy by leveraging extensive computational resources, without much concern for cost\cite{schwartz2020green}. This method relies on high-performance computing, large datasets, and complex models to achieve superior precision and performance, essentially obtaining better results by increasing computational investment.
    
    \item \textbf{Green AI} refers to artificial intelligence that achieves desirable results while being aware of computational expenses, encouraging the reduction of resource usage while maintaining efficiency\cite{schwartz2020green}. This means that, for designing and deploying {AI} models, there is a greater focus on reducing energy consumption and carbon emissions.
\end{enumerate}

\subsection{Proposed generalized red and green {AI}}\label{subsec_GGAI_GRAI}
The above {RAI} ang {GAI} distinctions have been strongly coupled to computation-, energy-, and carbon-related metrics. However, the notion of ``green'' has been defined more generally as multi-dimensional. In particular, \cite{J_ISJ_GBD_wu2016big} formalized ``green'' over multiple sustainability dimensions, including Environmental (E), Economic (Ec), Social (S), and Technical/Operational (T) criteria. The formal union- and intersection-based definitions reflecting the relations of sustainable and friendly are (\ref{equation_green_def1}) in \cite{J_ISJ_GBD_wu2016big}  and (\ref{equation_green_def2}) proposed here, provided in the Appendix.

Motivated by these general definitions, we propose the concepts of generalized green {AI} ({GGAI}) and generalized red {AI} ({GRAI}).
\begin{enumerate}
    \item \textbf{Generalized red AI (GRAI)} refers to approaches, systems, applications, and uses of artificial intelligence that move away from the goals defined by (\ref{equation_green_def1}) or (\ref{equation_green_def2}).
    
    \item \textbf{Generalized green AI (GGAI)} refers to approaches, systems, applications, and uses of artificial intelligence that move toward the goals defined by (\ref{equation_green_def1}) or (\ref{equation_green_def2}).
\end{enumerate}
Importantly, {GGAI}/{GRAI} evaluate not only the sustainability of \emph{AI services} (provided by developers/operators), but also the sustainability of \emph{AI uses} (downstream deployment and user-side consequences).

\subsection{Comparison and practical interpretation}\label{subsec_diff_GGAI_GRAI}
Table~\ref{tab:gai_vs_ggai} summarizes the conceptual differences between the traditional ({RAI}/{GAI}) and generalized ({GRAI}/{GGAI}) viewpoints.

\begin{table*}[pt]
\caption{Comparison between traditional (RAI/GAI) and generalized (GRAI/GGAI) concepts.}
\label{tab:gai_vs_ggai}
\centering
\small
\setlength{\tabcolsep}{4pt}
\renewcommand{\arraystretch}{1.15}
\begin{tabularx}{\textwidth}{p{0.12\textwidth} p{0.20\textwidth} X p{0.25\textwidth}}
\toprule
\textbf{Concept} & \textbf{Scope} & \textbf{Primary objective} & \textbf{Typical evaluation cues} \\
\midrule
RAI  & Model training/design & Accuracy via brute-force scaling & SOTA accuracy; larger models/data/compute \\
GAI  & Model training/design & Similar utility with less compute/power & FLOPs, time, energy, CO$_2$ emissions \\
GRAI & System / application / use & Away from environmental, economic, social, and technical/operational sustainability goals & Net degradation in one or more dimensions (may appear efficient locally) \\
GGAI & System / application / use & Toward environmental, economic, social, and technical/operational sustainability goals & Net improvement in one or more dimensions under a stated boundary \\
\bottomrule
\end{tabularx}
\end{table*}

\noindent\textbf{Examples (GGAI versus GRAI)}
\emph{GGAI instance:} training or serving an LLM under carbon-aware scheduling and workload shifting, with transparent reporting of both energy/carbon and water impacts (on-site cooling and power-related water), while avoiding disproportionate burden on water-stressed communities; the net effect is a measurable improvement along at least one of Environmental-Economic-Social-Technical/Operational.
\emph{GRAI instance:} scaling training/inference demand without accounting for regional grid intensity, basin water scarcity, and hardware turnover, leading to net deterioration in E (carbon/water/e-waste) and/or S (inequity) even if short-term efficiency metrics appear improved.

\subsection{Case study and how the framework guides engineering and policy}\label{subsec_case_GGAI_GRAI}

\textit{Illustrative case study (service side $\rightarrow$ use side):}
We consider training (or serving) a GPT-3-class large language model under carbon-aware scheduling and regional workload shifting, relative to a stated baseline (e.g., always-on, region-agnostic operation). Compared with always-on, region-agnostic scheduling, carbon-aware shifting can reduce on the order of 20\% CO$_2$e (such as $\approx$18--25\% reported under no-slack settings with the same completion-time constraint)\cite{10.1145/3663529.3663862}, depending on grid carbon-intensity variability and workload flexibility. On the service side, the operator shifts non-urgent training and batch inference to lower-carbon hours/regions, discloses electricity/carbon and water impacts (including on-site cooling and power-related water), and enforces constraints that prevent burden shifting (such as caps on water-stress-weighted water use in high-stress basins). On the use side, the demand is shaped by matching service quality to user needs: conditional-compute serving (such as mixture-of-experts routing \cite{J_TKDE_CaiWL_n7_2025} or compressed variants) allocates lower-compute paths when high fidelity is unnecessary, thereby reducing avoidable inference expenditure while meeting user-facing requirements. Under the stated boundary, if these interventions produce a verifiable net reduction in key footprint indicators (at minimum electricity/carbon and water, with hardware turnover as a lifecycle check) without violating explicit service constraints (such as latency/completion-time targets and cost caps) and without increasing inequitable externalized burdens, the instance is classified as generalized green artificial intelligence; otherwise it is classified as generalized red artificial intelligence. In particular, scaling training and inference without accounting for grid intensity, basin water scarcity, and hardware refresh can yield net deterioration in environmental and/or social outcomes even when short-term utilization or cost metrics appear favorable.

\textit{How to use the framework in practice.}
Given an artificial intelligence system, GGAI/GRAI can be operationalized as the following design-and-governance procedure:
\begin{enumerate}
    \item \textbf{Specify the assessment boundary and stakeholders:} separate the \emph{service} boundary (developer/operator side) from the \emph{use} boundary (deployment and user side), and state whose outcomes are evaluated.
    \item \textbf{Select a minimal indicator set across the four dimensions:} choose a small set of environmental (such as electricity/carbon, water, materials/e-waste), economic (such as total cost of ownership, resource productivity), social (such as equity and externalized burdens), and technical/operational indicators (such as reliability/latency under realistic workloads), consistent with the boundary.
    \item \textbf{Assess net change under constraints:} evaluate net improvement vs.\ degradation across the indicators relative to a baseline, accounting for lifecycle stages and region-specific constraints when applicable.
    \item \textbf{Classify and prioritize interventions:} if the net direction is toward sustainability goals, classify as generalized green; otherwise classify as generalized red and prioritize interventions targeting the limiting dimension(s), such as carbon-/water-aware scheduling, cooling choices, hardware lifetime extension, demand shaping, and governance constraints.
\end{enumerate}

\textit{Engineering design and policy implications.}
The generalized concepts make sustainability a boundary-aware design objective rather than an after-the-fact metric. On the engineering side, they motivate regional/temporal workload placement, joint optimization of carbon and water impacts, conditional-compute serving, and lifecycle-aware hardware refresh policies. On the governance side, they provide a reusable \emph{reporting and benchmarking interface} across service and use boundaries---a \emph{minimum disclosure set} including electricity use, grid carbon intensity (and resulting CO$_2$e), water-use effectiveness (WUE) / cooling water use, and hardware refresh or lifetime metrics---together with scenario-specific thresholds for carbon/water/e-waste that prevent burden shifting while preserving essential utility.

}

\section{Big Data and Information and Communication Technologies: Red and Green}\label{sec_BD_ICT_RG}
{
\color{black} 

\subsection{Motivation}
AI could be considered as a subset of information and communication technologies (ICT), while ICT are critical to artificial intelligence sustainability because they constitute the physical and operational substrate on which data are collected, stored, transmitted, and processed. In practice, the dominant environmental and resource footprints of modern artificial intelligence systems are often realized not only inside model training loops, but also through ICT-controlled decisions such as where and when workloads run (data centers and edge facilities), how data pipelines are scheduled (storage, replication, and transfer), and how frequently network and computing hardware is refreshed. Therefore, sustainability assessment cannot be confined to model-level efficiency alone, but it must extend to the coupled big-data and information and communication technologies stack and must distinguish impacts arising from \emph{service provision} (operators) versus \emph{service use} (downstream demand).
\noindent\emph{Illustration 1 (data-center scheduling policy):}
A carbon-aware scheduling policy can shift non-urgent training and batch inference to lower-carbon hours or regions and reduce peak-power stress, whereas ``always-on'' peak-hour processing and over-provisioned redundancy can increase electricity consumption and associated emissions even if per-request latency improves.

\noindent\emph{Illustration 2 (network hardware refresh cycles):}
Frequent hardware refresh cycles in networks and data centers may improve nominal performance-per-watt, yet can increase embodied impacts and electronic waste through accelerated equipment turnover; conversely, lifetime extension, reuse, and repair programs can reduce life-cycle burdens while maintaining required service levels.

\subsection{Relevant Definitions to Red and Green}\label{subsec_def_RG}
Prior works have explored the crises and opportunities presented by Big Data (BD) and Information and Communication Technologies (ICT) from a sustainability perspective \cite{J_ISJ_GBD_wu2016big,J_ISJ_BDG_wu2016big,wu2018ICTSDG}. However, specific terminologies for the "red" side of BD and ICT have not been fully formalized. Inspired by the {GAI}/{RAI} classification \cite{schwartz2020green} and our generalized definitions of {GGAI}/{GRAI} in Section \ref{sec_GGAI_GRAI}, we propose the following red concepts for BD and ICT based on the Environmental-Economic-Social-Technical/Operational
 framework (see Equations (\ref{equation_green_def1}) and (\ref{equation_green_def2})).

\subsection{Proposed Red Concepts}\label{subsec_red_BDICT}
\begin{enumerate}
    \item \textbf{Red Big Data (RBD)} refers to Big Data approaches, systems, applications, and uses that deviate from the sustainability goals defined by  (\ref{equation_green_def1}) or (\ref{equation_green_def2}). These may be considered excessive to the world, leading to negative impacts on Environmental (E), Economic (Ec), Social (S), or Technical/operational (T) outcomes. 
    
    \item \textbf{Red information and communication technologies (RICT)} refers to information and communication technologies approaches, systems, applications, and uses that deviate from the sustainability goals defined by  (\ref{equation_green_def1}) or (\ref{equation_green_def2}).  These may be considered excessive to the world, leading to negative impacts on Environmental (E), Economic (Ec), Social (S), or Technical/Operational (T) outcomes.
\end{enumerate}

Not only should {RBD} and {RICT} consider the sustainability of the provided BD or ICT services, but they should also assess the sustainability of the uses of these services. Specifically, {RBD} and {RICT} need to evaluate whether  the provided {BD} or {ICT} services as well as the {BD} or {ICT} uses are not green and sustainable and excessive to the world, thus exceeding acceptable thresholds in terms of Environmental-Economic-Social-Technical/Operational.

\subsection{Clarified Green Concepts}\label{subsec_green_BDICT}
In contrast, the concept of Green Big Data (GBD) has been proposed in previous works \cite{J_ISJ_GBD_wu2016big,J_ISJ_BDG_wu2016big}, which aligns with the green definition provided in  (\ref{equation_green_def1}).  In 2016 when \cite{J_ISJ_GBD_wu2016big,J_ISJ_BDG_wu2016big} were published, the significant wave of artificial intelligence represented by deep learning just began to rise, but large language models ({LLMs}) that require enormous computing power did not start until around 2019\cite{J_OAI_radford2019language} and 2020\cite{J_NIPS_brown2020language}. In 2020, the term green artificial intelligence ({AI}) emerged, in fact, accompanied by the emergence of large language models ({LLMs}) \cite{schwartz2020green}. The terms green {AI} and our proposed generalized green {AI} are truly highly meaningful only if they come with large-scale or big data-based systems. In other words, when big data are available, such as for large language models ({LLMs}), green {AI} and generalized green AI would converge to or can be considered as a subset of green big data. The issues of green big data analyzed in \cite{J_ISJ_GBD_wu2016big,J_ISJ_BDG_wu2016big} have become more and more important and difficult to be solved for the world in the current era of large language models and generative {AI}.

The Green Big Data and Green information and communication technologies concepts extend the green definitions and are clarified as follows:
\begin{enumerate}
    \item \textbf{Green Big Data (GBD)} refers to Big Data systems, applications, and approaches that move toward the sustainability goals defined by either $\mathrm{Green}_{\cup}$ or $\mathrm{Green}_{\cap}$, demonstrably improving one or more (preferably all) of the Environmental-Economic-Social-Technical/Operational
 dimensions.
    
    \item \textbf{Green information and communication technologies (GICT)} refers to ICT systems, applications, and approaches that move toward the sustainability goals defined by $\mathrm{Green}_{\cup}$ or $\mathrm{Green}_{\cap}$, demonstrating measurable improvements in one or more (preferably all) of the Environmental-Economic-Social-Technical/Operational dimensions.
\end{enumerate}

Figure~\ref{fig:ai_bd_ict_loop} summarizes the coupled AI--BD--ICT stacks and how sustainability impacts and constraints feed back into system design and use.


\begin{sidewaysfigure*}[pt]
\centering
\resizebox{0.98\textheight}{!}{%
\begin{tikzpicture}[
    font=\small,
    node distance=16mm,
    box/.style={draw, rounded corners, align=center, inner sep=6pt, minimum width=40mm},
    sbox/.style={draw, rounded corners, align=center, inner sep=5pt},
    arr/.style={-Latex, thick},
    darr/.style={-Latex, thick, dashed}
]

\node[box] (ai) {Artificial intelligence\\(models, training, inference)};
\node[box, below left=18mm and 30mm of ai] (bd) {Big data\\(collection, storage, curation, pipelines)};
\node[box, below right=18mm and 30mm of ai] (ict) {Information and communication technologies\\(data centers, networks, hardware, cooling)};

\draw[arr] (ai) -- node[left,  align=center] {\footnotesize consumes/produces\\data \& compute} (bd);
\draw[arr] (bd) -- node[right, align=center] {\footnotesize enables\\training \& updates} (ai);

\draw[arr] (bd) -- node[below, align=center] {\footnotesize depends on\\storage/transfer} (ict);
\draw[arr] (ict) -- node[above, align=center] {\footnotesize provides\\capacity \& connectivity} (bd);

\draw[arr] (ai) -- node[above right, align=center] {\footnotesize drives\\workloads} (ict);
\draw[arr] (ict) -- node[above left, align=center] {\footnotesize constrains\\performance/availability} (ai);

\node[sbox, above=10mm of ai, minimum width=140mm] (sus) {Sustainability impacts and constraints\\
(environmental: electricity/carbon/water/materials; economic: cost/productivity; social: equity/externalities; technical/operational: reliability/efficiency)};

\draw[darr] (sus.south west) .. controls +(0,-10mm) and +(-10mm,10mm) ..
    node[left, align=center] {\footnotesize metrics\\\& governance} (bd.north);
\draw[darr] (sus.south) .. controls +(0,-12mm) and +(0,12mm) ..
    node[right, align=center] {\footnotesize scheduling\\\& policies} (ai.north);
\draw[darr] (sus.south east) .. controls +(0,-10mm) and +(10mm,10mm) ..
    node[right, align=center] {\footnotesize infrastructure\\choices} (ict.north);

\draw[thick, dotted] ($(bd.west)+(-18mm,0)$) -- ($(ict.east)+(18mm,0)$);
\node[font=\footnotesize, above left=1mm and -2mm of bd.west] {Service provision side};
\node[font=\footnotesize, above right=1mm and -2mm of ict.east] {Use side (deployment and demand)};

\end{tikzpicture}%
}
\caption{Coupled sustainability loop between artificial intelligence, big data, and information and communication technologies, with feedback through footprint indicators and governance across service and use boundaries.}
\label{fig:ai_bd_ict_loop}
\end{sidewaysfigure*}
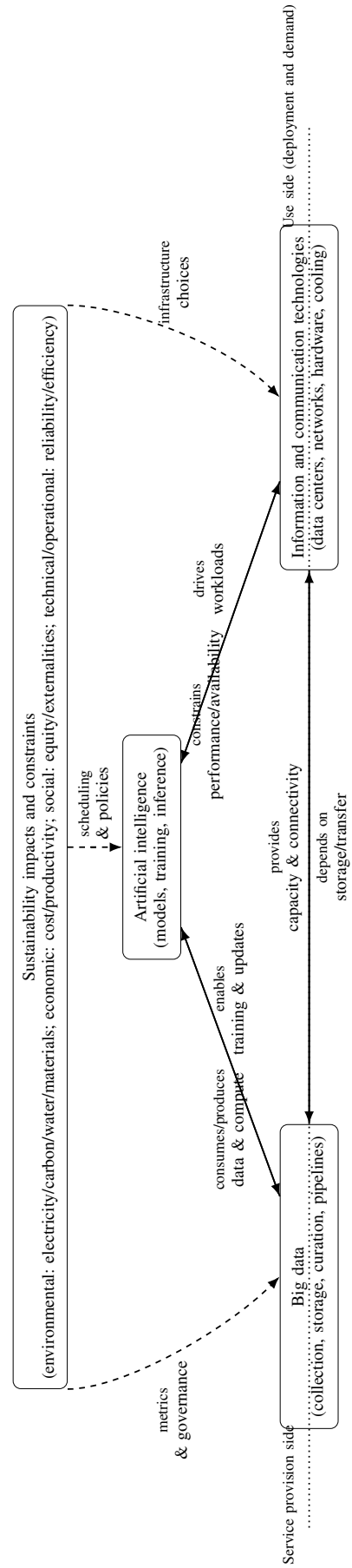

\subsection{Examples}\label{subsec_ex_BDICT}
\begin{enumerate}
    \item \textbf{RBD Examples:} Massive, ungoverned data storage without recent uses; indiscriminate data retention; processing data at peak-carbon hours; excessive data usage without regard for Environmental-Economic-Social-Technical/Operational dimensions.
    \item \textbf{RICT Examples:} Over-provisioned information and communication technologies systems without adequate use cases; rapid hardware refresh cycles generating e-waste; information and communication technologies deployments that consume excessive resources (energy/materials) relative to Environmental-Economic-Social-Technical/Operational
 sustainability.
    \item \textbf{GBD Examples:} Dataset curation and deduplication that reduce computing needs; privacy-preserving, energy-efficient data processing techniques; sustainable batch scheduling that aligns with SDGs.
    \item \textbf{GICT Examples:} Carbon- and latency-aware workload placement; extending hardware lifetime and promoting reuse; efficient information and communication technologies processing that supports SDGs and minimizes resource consumption.
\end{enumerate}

\subsection{Further Generalizations to Electronic and Electrical Engineering and Digital Technologies}\label{subsec_gen_elec_DT}
Similar to the concepts of RBD, RICT, GBD, and GICT, we may extend these concepts to other domains such as electronic and electrical engineering (EEE) and digital technologies (DT). The green and red definitions can be generalized as follows:
\begin{enumerate}
    \item \textbf{Green Electronic and Electrical Engineering (GEEE)}: Refers to approaches, systems, applications, and uses in EEE that demonstrably improve sustainability across Environmental-Economic-Social-Technical/Operational dimensions.
    
    \item \textbf{Green Digital Technologies (GDT)}: Refers to approaches, systems, applications, and uses in digital technologies that promote sustainability and reduce environmental impact in terms of Environmental-Economic-Social-Technical/Operational.
\end{enumerate}

Similarly, the red concepts can be defined as:
\begin{enumerate}
    \item \textbf{Red Electronic and Electrical Engineering (REEE)}: refer to approaches, systems, applications, and uses of electronic and electrical engineering that move away from the goals defined by (\ref{equation_green_def1}) or (\ref{equation_green_def2}). These may be considered excessive to the world, leading to negative impacts on Environmental-Economic-Social-Technical/Operational outcomes.
    
    \item \textbf{Red Digital Technologies (RDT)}: refer to  approaches, systems, applications, and uses of digital technologies that move away from the goals defined by (\ref{equation_green_def1}) or (\ref{equation_green_def2}). These may be considered excessive to the world, leading to negative impacts on Environmental-Economic-Social-Technical/Operational outcomes.
\end{enumerate}
These generalizations reflect the same principles as those for RBD, RICT, GBD, and GICT, and their practical applications can similarly be assessed for sustainability impacts.

}

\section{Current and future risks and possible countermeasures to generalized red AI, red BD, and red ICT in the era of large language models and generative AI}\label{sec_cr_Red}

The increasing significance of crises associated with generalized red AI has been highlighted in Section \ref{sec_crisis_AI}. This issue has become even more pressing with the advent of large language models (LLMs) and generative AI \cite{J_Na_jo2023promise}. Furthermore, as discussed in Section \ref{sec_BD_ICT_RG}, when large-scale systems and big data are involved, generalized red AI can be considered a subset of red big data (RBD). Beyond GRAI, RBD also includes non-AI computations. Meanwhile, information and communication technologies constitutes the underlying infrastructure for both computing and communications, and its evolution increasingly couples AI services and uses with network-wide energy/resource footprints. In this section, we explore countermeasures aimed at mitigating these growing risks under the broad Environmental-Economic-Social-Technical/Operational-oriented perspective.

\medskip
\noindent\emph{Countermeasure taxonomy.}
Consistent with the service--use boundary in Section~\ref{sec_GGAI_GRAI} and the Environmental-Economic-Social-Technical/Operational framework, the countermeasures are organized into four tiers: 
\begin{enumerate}
    \item Metric-level reform for information and communication technologies/data-center evaluation (service-side accountability across Environmental-Economic-Social-Technical/Operational),
    \item Information and communication technologies design-principle revision for provisioning and demand shaping (joint service-and-use optimization), 
    \item LLM development practices (data, model, training, and scaling) that improve performance-per-resource while controlling social risks,
    \item Energy-infrastructure transition (renewables/nuclear) that shifts the underlying supply constraints affecting both services and uses.
\end{enumerate}

Table~\ref{tab:roadmap_countermeasures} provides a time-phased roadmap (near-, medium-, and long-term) that links countermeasures to both service provision and downstream use, spanning environmental, economic, social, and technical/operational impacts.
\begin{table*}[!t]
\caption{Roadmap of countermeasures (near-, medium-, and long-term) for generalized red artificial intelligence, red big data, and red information and communication technologies across service provision and downstream use.}
\label{tab:roadmap_countermeasures}
\centering
\small
\setlength{\tabcolsep}{4pt}
\renewcommand{\arraystretch}{1.15}
\begin{tabularx}{\textwidth}{p{0.16\textwidth} X X X}
\toprule
\textbf{Category} & \textbf{Near-term} & \textbf{Medium-term} & \textbf{Long-term} \\
\midrule
\textbf{Metrics \& accountability} &
Standardize footprint disclosure (electricity/carbon/water/e-waste) for training and inference; report impacts for both services and uses. &
Audit-ready lifecycle accounting with region/time awareness; introduce threshold-based compliance and equity safeguards. &
Institutionalize multidimensional sustainability labeling and procurement standards; enable longitudinal benchmarking. \\
\textbf{System design \& demand shaping} &
Deploy carbon-aware scheduling for non-urgent workloads; avoid peak-hour batch pipelines; reduce over-provisioned redundancy. &
Capacity planning guided by demand distributions (not peak-only design); introduce service-tiering and user-side incentives. &
Sustainability-by-design architectures for networks and data centers aligned with societal value and constraints. \\
\textbf{Large-model development} &
Data curation/deduplication; selective invocation; pruning/distillation where acceptable. &
Training/inference co-optimization under resource constraints; systematic multi-objective evaluation beyond accuracy. &
Efficiency-first paradigms and incentives; shift away from brute-force scaling as the default development mode. \\
\textbf{Energy \& infrastructure transition} &
Increase low-carbon electricity use where available; improve cooling and water stewardship; extend hardware lifetime and reuse. &
Integrate clean energy with storage and flexible loads; reduce embodied impacts via circular hardware supply chains. &
Deep decarbonization of power/material/manufacturing with robust water and waste governance at scale. \\
\bottomrule
\end{tabularx}
\end{table*}

\subsection{Reconsidering design green measures}
The designs of existing data centers and information and communication technologies systems have largely focused on improving energy efficiency. However, energy-centric measures alone are insufficient to characterize sustainability risks in the LLM era. More comprehensive metrics such as \emph{effective energy efficiency} and \emph{effective resource efficiency} should be prioritized, considering the full lifecycle of both data and model consumption \cite{J_ISJ_GBD_wu2016big}. In addition, consistent with the generalized concepts in Sections \ref{sec_GGAI_GRAI} and \ref{sec_BD_ICT_RG}, it is preferable to evaluate not only the sustainability of provided services, but also the sustainability of downstream uses, i.e., whether the combined service-and-use pattern is green/sustainable rather than excessive with respect to Environmental-Economic-Social-Technical/Operational.

\medskip
\noindent\emph{Environmental-Economic-Social-Technical/Operational linkage:}
These measures primarily target \emph{E} (energy, carbon, water, and materials over the lifecycle) and \emph{Ec} (total cost of ownership and resource productivity), while also supporting \emph{T} (system-level efficiency under realistic workloads) and \emph{S} (avoiding externalized burdens through service--use evaluation).

\subsection{Rethinking design motivations}
Although the fifth generation (5G) wireless communications have achieved notable energy-efficiency improvements relative to 4G, 5G base stations often have smaller coverage and thus require denser deployments, which may increase system-level power demand. A 2023 study on 5G networks in China identified a ``carbon efficiency trap'' caused by the mismatch between the spatiotemporal distribution of cellular traffic and the energy consumption of networks \cite{J_NatS_li2023carbon}. In parallel, several studies have begun to consider the involvement of LLMs and generative AI in future communications technologies \cite{M_ComMag_bariah2024large,M_ComMag_AMaatouk_2024,zhou2024LLMtelecommm}, which may introduce new risks in GRAI and RBD as well as RICT.

The following issues could be reviewed in future system designs regarding data volume, transfer rates, and service provisioning, from both the service-provider side and the user side:
\begin{enumerate}
    \item \emph{Service provider side (service provisioning):}
    Historically, information and communication technologies designs, especially in wireless communications from 1G to 6G, have focused primarily on maximizing peak data volumes and peak transfer rates. However, beyond 5G, such peak-driven design principles should be revisited. A recent study \cite{J_TSG_zou2022electric} pointed out that the probability of very high data rates decays rapidly, implying that systems designed mainly for extreme peaks can suffer from low utilization and poor system-level efficiency. Similar views apply to other stringent requirements (such as ultra-low latency and extreme mobility). To understand these issues more closely, it is necessary to investigate empirical distributions of design requirements across temporal, spatial, and user dimensions. A practical direction is to adopt probabilistic or tiered provisioning strategies: designing for the dominant mass of requirements while treating extreme demands as low-probability cases with targeted mechanisms. In addition, providers should improve AI/LLM efficiencies in energy and resources for both internal information and communication technologies operations and provided service pipelines, such as via carbon-/resource-aware scheduling and workload management.
    
    \item \emph{User side (use patterns and demand shaping):}
    System designs should reconsider whether all user demands are treated as identical in sustainability implications. Over-provisioning (such as persistently delivering very high access rates to all users regardless of actual needs) can induce avoidable energy/resource waste. For example, nowadays many modern cities provide services of 300 - 500 Megabytes per second (Mbps) data rate optical accesses for most of users. However, 100 Mbps data rate optical accesses are actually sufficient for many users, and those over-provided data rate optical accesses are provided at sacrifice of wasted energy and resources. Those over-provided data rate issues have also been widely available for wireless accesses.
    A more sustainable approach is to differentiate service tiers and encourage demand shaping, where non-critical high-throughput consumption is reduced without harming essential societal functions. Moreover, AI/LLM service demands are also highly heterogeneous across time, locations, and user groups, and thus system designs should avoid assuming that maximum AI/LLM service must be provided uniformly across all periods, all users, and all locations.
\end{enumerate}

\medskip
\noindent\emph{Environmental-Economic-Social-Technical/Operational linkage:}
Probabilistic/tiered provisioning and carbon-aware operations directly improve \emph{E} (lower network-wide energy and emissions) and \emph{Ec} (reduced over-provisioning and OPEX/CAPEX pressure), preserve \emph{T} by meeting the dominant QoS distribution rather than rare extremes, and can improve \emph{S} via more equitable and transparent service tiers and demand shaping.

\subsection{Reconsidering the development of large models}
\subsubsection{Data-Model efficiency methods}
Both dataset choices and model/algorithm design jointly determine the computational workload and lifecycle footprint of large-model development. On the data side, energy efficiency can vary significantly with dataset modification and curation strategies \cite{10092659}, and data-centric practices (e.g., deduplication, filtering, and task-aligned data selection) can reduce unnecessary training/inference cycles and associated emissions while preserving utility \cite{verdecchia2022data}. On the model side, different architectures and training pipelines may consume substantially different resources to achieve similar task performance \cite{georgiou2022green}. Therefore, ``efficiency-first'' methods should be prioritized, including lightweight adaptation instead of full retraining when feasible, as well as model compression and reuse techniques such as distillation, pruning, quantization, and transfer learning \cite{chen2023survey,mcdonald2022great,hadi2023survey,kasneci2023chatgpt,weidinger2022taxonomy}. These methods aim to reduce redundant computation and memory movement, lowering electricity demand and carbon/water footprints with controlled accuracy loss. Importantly, greener configurations are not necessarily inferior in model quality; empirical evidence suggests that higher footprint does not imply better model performance, motivating systematic evaluation of quality--footprint trade-offs rather than default brute-force scaling \cite{everman2023evaluating}.

\medskip
\noindent\emph{Environmental-Economic-Social-Technical/Operational linkage:}

\begin{enumerate}
    \item Data Side:
    Dataset optimization mainly improves \emph{E} and \emph{Ec} by reducing redundant collection, storage, and training signals, supports \emph{T} by improving data quality and lowering pipeline overhead, and can strengthen \emph{S} when representativeness, privacy, and downstream harms are explicitly considered.
    \item Model side:
    Model and algorithm refinement primarily targets \emph{E} and \emph{Ec} through reduced computation and energy cost per delivered utility, and supports \emph{T} by meeting reliability/latency requirements under resource constraints; it should be coupled with explicit assessment of \emph{S} (such as fairness, bias, and misuse risk) to avoid shifting impacts from environmental burdens to societal harms.
\end{enumerate}

\subsubsection{Adjusting training processes}
Beyond model structure, training processes can also be redesigned to avoid unnecessary high-cost computation. For example, selectively applying AI/LLM computation only where it is needed (rather than uniformly throughout pipelines) may reduce energy consumption and carbon and water footprints with acceptable trade-offs in time \cite{li2023making,schwartz2020green}.

\medskip
\noindent\emph{Environmental-Economic-Social-Technical/Operational linkage:}
Selective or stage-wise training reduces resource-intensive computation most directly in \emph{E} and \emph{Ec}, can improve \emph{T} by avoiding unnecessary peak provisioning, and requires checks on \emph{S} to ensure that efficiency gains do not come from unsafe shortcuts.

\subsubsection{Questioning scaling laws}
Scaling laws, such as those proposed by Kaplan et al. \cite{kaplan2020scaling}, suggest diminishing returns as model size increases: while performance improves with scale, the marginal improvement becomes smaller. Increasing dataset size can improve performance for a given model size, indicating that data scaling is crucial for large models to realize their potential. However, Bender et al. \cite{bender2021dangers} cautioned that simply making models larger does not necessarily yield better generalization or responsible outcomes, motivating more responsible development that integrates sustainability considerations. Kumar et al. \cite{kumar2024scaling} further refined scaling laws by introducing precision scaling, showing that numerical precision choices can reduce training costs compared with brute-force scaling. Collectively, these results support shifting from unsustainable scaling to efficiency-oriented scaling that explicitly accounts for sustainability, fairness, and bias mitigation.

\medskip
\noindent\emph{Environmental-Economic-Social-Technical/Operational linkage:}
Efficiency-oriented scaling (such as precision scaling and responsible data scaling) addresses \emph{E/Ec} by lowering marginal resource growth, supports \emph{T} via better performance-per-watt under deployment constraints, and aligns with \emph{S} by enabling governance targets (fairness and harm mitigation) beyond scale-only optimization.

\subsection{Increasing use of low-carbon energy and complementary demand-side controls}\label{subsec_energy_transition}
The electricity demand of artificial intelligence, big data, and information and communication technologies increasingly couples digital growth to power-system emissions. Expanding low-carbon electricity supply (renewables and, in some regions, nuclear) can reduce operational carbon intensity, but it does not remove the need for demand-side controls because intermittency, transmission constraints, and local resource limits (such as water availability for cooling) can shift burdens across time and regions.

From an AI/BD/ICT perspective, energy transition should be implemented together with system-specific interventions: 
\begin{enumerate}
    \item treating non-urgent training and batch processing as flexible loads and shifting them to lower-carbon hours/regions, 
    \item  incorporating grid constraints and water-stress signals into workload placement and data-pipeline scheduling, 
    \item avoiding ``always-on'' over-provisioning that increases baseline consumption without proportional societal value. In parallel, nuclear energy can be a low-carbon option but faces non-trivial constraints in capital intensity and long-term radioactive waste governance \cite{J_PNE_MATHEW_2022104080}; renewable deployment faces intermittency and land/transmission constraints and may rely on critical materials with environmental and social externalities \cite{J_Eng_GH_en13112906_2020}. These trade-offs should be reflected explicitly in information and communication technologies and data-center planning rather than assumed away.
\end{enumerate}

\medskip
\noindent\emph{Policy and regulatory pathways.}
To operationalize the above transition, policy instruments can align incentives across both service provision and downstream use. Carbon pricing (or equivalent compliance mechanisms) can internalize emissions externalities and reward carbon-aware scheduling and siting choices; energy/carbon labeling and mandatory disclosure can enable comparability across providers and services; and procurement standards (such as clean-energy procurement requirements for large deployments) can create enforceable demand-side pressure toward low-carbon operation rather than purely performance-driven scaling.

\medskip
\noindent\emph{Environmental-Economic-Social-Technical/Operational linkage.}
Energy-source transition primarily targets \emph{E} (decarbonization and reduced local environmental stress) but introduces non-trivial \emph{Ec} (investment and grid/storage costs), \emph{T} (reliability under intermittency and operational constraints), and \emph{S} considerations (land use, material supply chains, and long-term risk governance); hence it must be coupled with AI/BD/ICT-specific workload flexibility and governance to avoid burden shifting.

\section{Conclusion}
{
\color{black} 
This paper has proposed a unified theoretical framework to assess and govern the sustainability implications of artificial intelligence (AI), big data (BD), and information and communication technologies (ICT) by introducing generalized green/red concepts within a multi-dimensional sustainability framework spanning \emph{environmental, economic, social, and technical/operational} dimensions. Building on this framework, we have clarified how generalized red practices can emerge not only from model-level scaling but also from data pipelines, infrastructure provisioning, and downstream uses, and we have outlined actionable countermeasures that connect engineering levers (such as green measure re-considerations, design motivation re-considerations, carbon-aware workload schedulings, and hardware lifecycle extensions) to governance levers. The expected policy impact is to provide a structured basis for translating abstract sustainability goals into operational requirements, including comparable disclosure, labeling, and procurement criteria, so that incentives for service provision and downstream use can be aligned toward generalized green directions rather than burden shifting. 

Despite establishing conceptual foundations for systemic sustainability assessment, this work has limitations. The proposed classifications (such as GGAI, GRAI, RBD, and RICT) are primarily conceptual constructions and currently lack large-scale empirical validation and scenario-specific quantitative metrics. Future research may develop additional scenario-specific measurable indicators (such as scenario-specific  multidimensional sustainability indices aligned with Environmental-Economic-Social-Technical/Operational) and conduct longitudinal case studies in hyperscalers, small and medium-sized enterprises and developing regions. Furthermore, while this paper has highlighted the risks related to electricity, water and e-waste, deeper modeling of indirect effects may be required, such as supply-chain emissions, embodied impacts, and the amplification of social harms (such as bias and inequitable access) through large-scale deployment.

Ultimately, transitioning from GRAI/RBD/RICT to GGAI/GBD/GICT introduces not only technical challenges but also governance challenges. It requires coordinated action among developers, policy makers, industries, and global institutions to align AI/BD/ICT services and their downstream uses with multidimensional sustainability objectives. Such progress depends on interdisciplinary collaboration across artificial intelligence, climate and energy systems science, and public policy to align technical design choices with measurable sustainability constraints. The overarching goal is to enable AI, BD, and ICT to serve as net enablers to a sustainable future  planet while reducing their potentials to become systemic obstacles under Environmental-Economic-Social-Technical/Operational constraints.

}

\appendix
The equations for the general green definitions are (\ref{equation_green_def1})\cite{J_ISJ_GBD_wu2016big} and (\ref{equation_green_def2}).
\begin{figure*}[pt]
\begin{eqnarray}\label{equation_green_def1}
\begin{gathered}
  {\text{Green}} = ({\text{Environmentally/Economically/Socially/Technically) Sustainable}} \hfill \\
   \cup {\text{ (Environmentally/Economically/Socially/Technically) Friendly}} \hfill \\ 
\end{gathered} 
\end{eqnarray}
\end{figure*}
\begin{figure*}
\begin{eqnarray}\label{equation_green_def2}
\begin{gathered}
  {\text{Green}} = ({\text{Environmentally/Economically/Socially/Technically) Sustainable}} \hfill \\
   \cap {\text{ (Environmentally/Economically/Socially/Technically) Friendly}} \hfill \\ 
\end{gathered}
\end{eqnarray}
\end{figure*}

\section*{Acknowlegement}
This work was supported in part by China Guangxi Science and Technology Plan Project under grant AD23026096, Chile CONICYT FONDECYT Regular under grant 1181809, and Chile CONICYT FONDEF under grant ID16I10466.

\bibliographystyle{IEEEtran}
\bibliography{IEEEabrv,Reference}

\end{document}